\documentclass[showpacs,preprintnumbers,superscriptaddress,amsmath,floatfix,amssymb,prd]{revtex4}
\usepackage{graphicx}
\usepackage[colorlinks=true]{hyperref}
\usepackage{amsfonts,amsmath,amsxtra}
\newcommand{\beq}[1]{\begin{eqnarray}\label{#1}}
\newcommand\eeq {\end{eqnarray}}
\newcommand\bqa {\begin{eqnarray}}
\newcommand\eqa {\end{eqnarray}}

\newcommand{\bear}{\begin{array}}
\newcommand{\enar}{\end{array}}

\newcommand{\R}{\mathbb{R}}
\newcommand{\C}{\mathbb{C}}
\newcommand{\K}{\mathcal{K}}

\newcommand{\Z}{\mathbb{Z}}

\newcommand{\N}{\mathbb{N}}

\begin{document}
\def\le{\langle}
\def\re{\rangle}
\def\dg{^{\dag}}
\def\K{{\cal K}}
\def\n{{\cal N}}
\font\maj=cmcsc10 \font\itdix=cmti10
\def\N{\mbox{I\hspace{-.15em}N}}
\def\H{\mbox{I\hspace{-.15em}H\hspace{-.15em}I}}
\def\1{\mbox{I\hspace{-.15em}1}}
\def\pN{\rm{I\hspace{-.1em}N}}
\def\Z{\mbox{Z\hspace{-.3em}Z}}
\def\pZ{{\rm Z\hspace{-.2em}Z}}
\def\R{{\rm I\hspace{-.15em}R}}
\def\pR{{\rm I\hspace{-.10em}R}}
\def\C{\hspace{3pt}{\rm l\hspace{-.47em}C}}
\def\Q{\mbox{l\hspace{-.47em}Q}}
\def\b{\begin{equation}}
\def\e{\end{equation}}
\def\bee{\begin{enumerate}}
\def\wt{\widetilde}

\title{Gravitational Merging as a Possible Source for the Cosmological Accelerating}
\author{Mehrdad Khanpour}
\email{m.khanpour@iauamol.ac.ir}
\affiliation{Faculty of Basic Sciences, Ayatollah Amoli Branch, Islamic Azad University, Amol, Iran}
\author{Ebrahim Yusofi}
\email{e.yusofi@iauamol.ac.ir (Corresponding author)}
\affiliation{Department of Physics, Faculty of Basic Sciences, Ayatollah Amoli Branch, Islamic Azad University, Amol, Iran}
\affiliation{School of Astronomy, Institute for Research in Fundamental Sciences(IPM), P. O. Box 19395-5531,Tehran, Iran}
\author{Bahman Khanpour}
\email{b.khanpour@stu.umz.ac.ir}
\affiliation{Department of Physics, Faculty of Basic Sciences, University of Mazandaran, P. O. Box 47416-95447, Babolsar, Iran}
\date{\today}
\begin{abstract}
   Gravitational merging (or clustering) of cosmic objects is regarded as a possible source of the extra-acceleration of the universe at large scale. The merging/clustering of cosmic objects introduces a correction term in the equation of state for the effective present cosmic fluid in the form of $P=w\rho+b\rho^2$. As a result, an alternative relation for the energy density includes over and under-dense regions is obtained that coincide with the conventional relation in the standard limit. By analogy with bubbles, the under-dense regions (voids) in the cosmic fluid is shown to provide the needed negative pressure. Invoking the observational constraint for the dark energy equation of state $w$, we show that the merging of voids will act as a possible source of extra-accelerating at large scale in comparison with non-merging cosmic gas.
\end{abstract}
\pacs{98.80.Jk, 95.36.+x, 98.80.Bp}

\maketitle
\section{Introduction and Motivation}
The inflationary models show acceleration in the early universe\cite{c1, c2, c3}. As of 1981 it is believed that in the far past time, the universe exponentially accelerated and then came out of this phase after a very short period. Before decoupling, the universe expansion was driven by radiation and particles, but after that driven only by matter, and hence it was thought that the expansion should be slowing down\cite{c4, c5, c6, c7}.\\
However, in 1998-99, contrary to the above picture, the accelerated expansion was confirmed by two groups from observations of supernova Ia explosion. WMAP and Planck recently confirmed the acceleration in the late universe\cite{c10, c11}. Naturally, a real model of cosmology requires a source of energy to drive acceleration in the present universe. In order to seek for a suitable source for dark energy, it is important to know the nature of the effective cosmic fluid in the large scales \cite{c4}. \\
Modern cosmology strongly depends on particle physics, and the role of scalar field is very significant in it\cite{c3}. So, cosmologists proposed scalar fields as a source of dark energy in several models. Some of the most important of them are Quintessence\cite{c12}, K-essence\cite{c13}, tachyon scalar\cite{c14}, Chaplygin gas\cite{c15, c151}, phantom scalar\cite{c16, c161} and generalized Chaplygin gas\cite{c17}, and so on. On the geometrical side, it has been proposed that the general relativity should be modified to account for dark energy, where some curvature functions are assumed to explain late acceleration. Some models in this regard include $f(R)$, and  $f(T)$ gravity\cite{c18, c19, c191, c192, c193}, where $R$ and $T$ are scalar curvature and scalar torsion, respectively. In this work, we try to explain the present accelerating universe by introducing some kind of real cosmic fluid (RCF).\\
The perfect cosmic fluid (PCF) for the whole universe with the equation of state (EOS)\cite{c4, c20},
\begin{equation}
\label{hard0}
P=w\rho,
\end{equation}
is usually invoked in most cosmological models, where $P$, $w$ and $\rho$ are the pressure, EOS parameter and energy density of the cosmic fluid, respectively. However in the real present universe, it can not accurately describe the physical properties of the cosmic fluid. \emph{One problem in the perfect cosmic gas law (1) is what happens to the local regions with higher densities in the large scales overview?} As the internal volume of the regions becomes smaller, their densities become higher.\\
In the ideal gas law it is assumed that the particles are point-like without any volumes and interactions, but in the real physical universe including clusters of galaxies, it is much better to be modeled by a more real gas system. In fact, there exist some phenomena occurring between cosmic objects in the $RCF$, deserve to be mentioned strictly. Merging/clustering of the cosmic objects is a common phenomenon at cosmological scale at present universe. Galaxies and galaxy clusters, and even voids are all always merging together to form large scale structure. We think that probably the merging/clustering of these objects may be responsible for some unsolved cosmological problems specially accelerating expansion at large scale universe. Of course, we note that objects of one type can only merge to each other. For example, a galaxy would merge with another galaxy and a void with another one.\\
\section{Cosmic Fluid with Merging Process}
The early universe cosmic fluid, as we know is so dilute that, as a whole we can consider it as an ideal gas. However, because of the importance of merging/clustering phenomena in the present state of the universe at cosmic scale, we would like to include its effects into the behavior of the cosmic fluid. Since the cosmic fluid is very dilute, the merging/clustering process can only occur between two objects of the same type, that we can write this as a simple \emph{merging process}; $2B = A$ \cite{c201}, where final object $A$ represents the merged one formed from two initial objects $B$. If $\varrho(=N/V)$ represents for the number density of the cosmic fluid then, based on the proposed merging process, we have $\varrho_{B}=\varrho-2\varrho_{A}$, where  $\varrho_{B}$ and $\varrho_{A}$ stand for the number densities of the cosmic fluid before and after the merging process, respectively. The equilibrium constant of the merging process is \cite{c201}
\begin{equation}
\label{hard00}
K(T)=\frac{\varrho_A}{(\varrho-2{\varrho_A})^2}\thickapprox \frac{\varrho_A}{\varrho^2},
\end{equation}
here $\varrho_A\ll \varrho$. The whole number density of the cosmic fluid due to merging is changed from $\varrho$ to $(\varrho- 2\varrho_A)+\varrho_A=\varrho-\varrho_A$. So we can write
$P=(\varrho-\varrho_A)k_{B}T$ or
\begin{equation}
\label{hard11}
\frac{P}{\varrho k_{B}T}=1-\frac{\varrho_A}{\varrho}=1-K(T)\varrho,
\end{equation}
here $K(T)>0$. Obviously, the volume of the merged galaxies are small fractions of the total cosmos volume, but the volume of the merged voids in contrast would be large fractions of the present cosmic web and their contributions will become larger. In addition, the energy density of merging regions is much higher than in other regions as depicted in FIG.1. Taking into account the contribution and effect of the volumes and interactions of these merged components, it also play an important role in the dynamics of the large scale of the cosmos. The main task of this article is to address this very common and important process between voids to solution of dark energy problem. The PCF $(\ref{hard0})$ does not account the merging process and its effects.\\
Let's suppose that in addition to presence of conventional cosmic fluid with the EOS $(\ref{hard0})$, there is another process in the cosmic fluid; their merging with each other. In the present universe, the galaxies, galaxy clusters and even voids are the possible natural candidates for these merging cosmic objects \cite{c202}. We thus see that these objects will contribute some portions of the volume of the whole cosmic fluid. This assumption, based on the real gas theory (accounting the merging process) by using $(\ref{hard11})$ and $E=\rho{V}=(1/w)Nk_{B}T$, leads to the following form of EOS\footnote{For example, in the follow-up to this topic, cosmic fluid with non-linear forms of EOS and its consequences have been discussed in some works such as \cite{c194, c195, c196, c197, c198}.}\cite{c20,c201},
\begin{equation}
\label{hard1}
P=w\rho+b\rho^2,
\end{equation}
where,
 \begin{equation}
\label{hard12}
b=-K(T)w^{2}/k_{B}T,
\end{equation}
is proportional to the equilibrium constant $K(T)$. $k_B$ and $T$ are the Boltzmann constant and temperature, respectively.\\
Note that the parameter $b$, taking into account the merging process is temperature dependent and negative. It seems that at lower temperatures the absolute values of it increases, meaning the merging of the cosmic objects coming more into play at lower and lower temperatures as expected from present (cold) universe. Thus the real cosmic gas $(\ref{hard1})$, is containing of two parts; the normal part $w\rho$, that occupies the larger contribution of the volume of the universe with low density, and the other part of the cosmic fluid, the merged one. These over-dense merging regions make interfaces with other under-dense non-merging parts and finally formed the cosmic web (see Fig. 1). \\
\begin{figure}[h]
\centering
\includegraphics[width=2.3in]{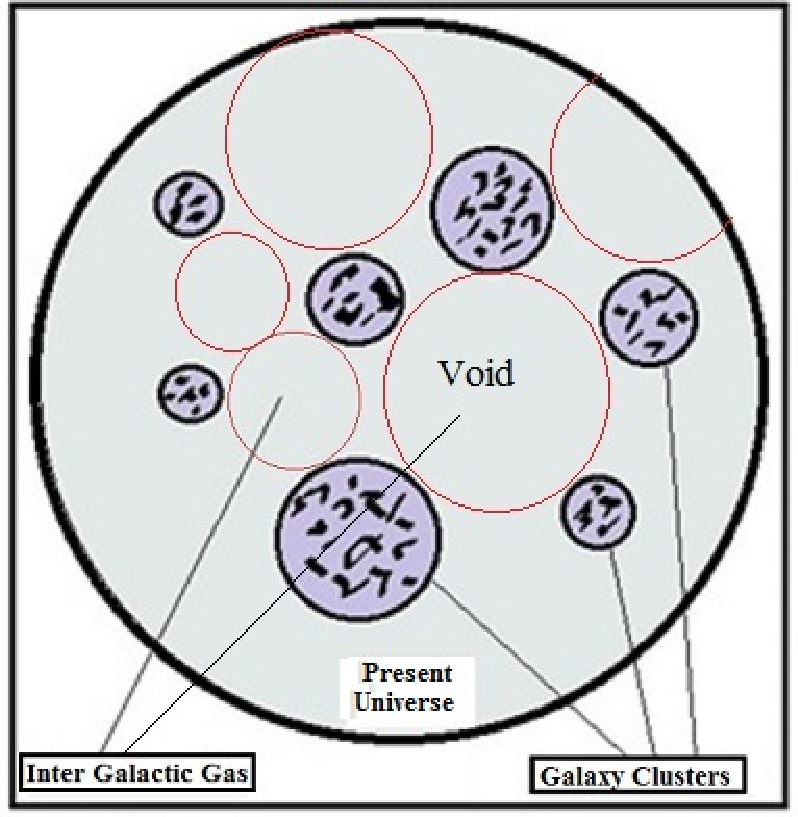}
\caption{Real Cosmic fluid includes some over (under)-dense merging regions\cite{c21}.}
\label{Fig1}
\end{figure}

\section{The Main Cosmological Relations}

\subsection{Energy Density in Terms of Scale Factor}
Energy density is one of the most important properties of the Universe. Energy density is defined as the energy within a region, divided by the volume of that region. Since the universe is homogeneous, it does not matter where the supposed regions are located. How can it be large? It depends on the epoch under consideration. In the early universe, before galaxy formation gets underway, such regions should contain many tiny particles. After the formation of galaxies the regions must contain many galaxies, and after the formation of galaxy clusters it must contain many galaxy clusters and voids in the form of the cosmic web \cite{c21}. \\
The Friedmann equations tells us the effect of gravity on the expansion rate. According to them, that effect is related to the scale factor itself, and the energy density \cite{c4, c5, c6, c7}:
 \begin{equation} \label{a1}
\frac{{{{\dot{a}}}^{2}}(t)}{{{a}^{2}}(t)}+\frac{k{{c}^{2}}}{R_{0}^{2}{{a}^{2}}(t)}=\frac{8\pi {{G}_{N}}}{3}\rho
 \end{equation}
and
\begin{equation} \label{a2}
\frac{\ddot{a}(t)}{a(t)}=-\frac{4\pi {{G}_{N}}}{{{c}^{2}}}\left( p+\frac{1}{3}\rho {{c}^{2}} \right)
 \end{equation}
where $a(t)$  represents the scale factor and $k$ is the curvature signature. These equations can be combined to get \cite{c5, c6},
\begin{equation} \label{a3}
\frac{d}{dt}\left( \rho {{c}^{2}}{{a}^{3}} \right)=-p\frac{d{{a}^{3}}}{dt}                                                                                                                  \end{equation}
We shall study the scaling behavior of the energy density in terms of scale factor. For the perfect fluid Eq.$(\ref{hard0})$, from (\ref{a3}) we have the standard result,
\begin{equation}
\label{hard6}
\rho=\rho_{0} a^{-3(1+w)}.
\end{equation}
It turns out that the expansion rate depends on where the energy density comes from. The part of the energy density that comes form particles are being attracted towards each other by the force of gravity which is what one would expect. Strangely though, the part of the energy density that comes from the cosmological constant (as a candidate) has the opposite effect and it increases the expansion rate of the large scale universe\cite{c21}. Instead of cosmological constant, as we will see later, the merging of over (under)-dense cosmic objects are now introduced to account for extra-accelerations observed in our universe.\\
To have a real cosmology with these two contrast parts of the energy density, we invoke our proposed EOS Eq.$(\ref{hard1})$.
Combination of two Eqs.$(\ref{a3})$ and $(\ref{hard1})$ gives:
\begin{equation} \label{a5}
\frac{d\rho }{(1+w )\rho + b{{\rho }^{2}}}=-3\frac{da}{a},                                                                                                                \end{equation}
which can be integrated to
\begin{equation} \label{a6}
\frac{\ln (\rho )}{(1+w )}-\frac{\ln (1+w + b\rho )}{(1+w )}=\ln (c{{a}^{-3}})                                                                                                               \end{equation}
or
\begin{equation} \label{a7}
\frac{\rho }{1+w + b\rho }=C{{a}^{-3(1+w )}}                                                                                                               \end{equation}
where $c$ and $C=\ln(c)$ are constants. The latter can be rearranged to get the following analytical formula:
\begin{equation} \label{a8}
\rho =\frac{C(1+w ){{a}^{-3(1+w )}}}{1- {B}{C}{{a}^{-3(1+w )}}}                                                                                                                \end{equation}
In order to find C we employ the usual conditions at which $\rho=\rho_0$ and $a=1$ when  $t=t_0$ , where subscript zero indicate the corresponding value at
the present time. This leads to
\begin{equation} \label{a9}
C=\frac{{{\rho }_{0}}}{1+w + b{{\rho }_{0}}}                                                                                                               \end{equation}
from which one obtains:
\begin{equation} \label{a10}
\rho =\frac{{{\rho }_{0}}(1+w ){{a}^{-3(1+w )}}}{1+w + b{{\rho }_{0}}- b{{\rho }_{0}}{{a}^{-3(1+w )}}}                                                                                                              \end{equation}
We can first rewrite Eq.$(\ref{a10})$ as
\begin{equation} \label{a11}
\rho =\frac{{{\rho }_{0}}{{a}^{-3(1+w )}}}{1+\alpha\left( 1-{{a}^{-3(1+w )}} \right)},
\end{equation}
where $\alpha =\frac{b{{\rho }_{0}}}{(1+w )}$. Since $\rho_0$ and consequently $\alpha$ are so tiny quantities, therefore the second term in the denominator which is showing merging effect in the energy density is less than one, i.e. $\alpha (1-a^{-3(1+w)}) < 1$. We can thus expand it to first order,
 \begin{equation}
\label{hard7}
\rho ={{\rho }_{0}}{{a}^{-3(1+w )}}\left( 1-\alpha\left( 1-{{a}^{-3(1+w )}} \right) \right).
\end{equation}
 to recast it into our desired final relation,
\begin{equation}
\label{hard71}
\rho_{r}=(1-\alpha){\rho }_{0}{{a}^{-3(1+w )}}+\alpha{\rho }_{0}{{a}^{-6(1+w )}}.
\end{equation}
Here, subscript $r$, represents the \emph{real} cosmic fluid (RCF).
\subsection{Evolution of the Scale Factor}
We will proceed to solve equation $(\ref{a1})$ for the time evolution of the scale factor $a(t)$. Here we assume that second term of the Eq.$(\ref{a11})$, as the first perturbation term for the energy density of the real universe.\\
In order to find the scale factor as a function of time, the equation $(\ref{a11})$ can now be combined with $(\ref{a1})$. This goal will be achieved by invoking perturbation theory by taking $\lambda$ as the perturbation coupling parameter and assuming that
\begin{equation}
\label{hard69}
a_r = a+\lambda{a_1},
\end{equation}
where $a_1$ is the perturbative part of the scale factor to be determined. We thus find (with another rationale assumption that $k=0$):
\begin{equation} \label{a13}
{{\left( \frac{{{{\dot{a}}}}}{{{a}}} \right)}^{2}}=\frac{8\pi {{G}_{N}}}{3}{{\rho }_{0}}a^{^{-3(1+w )}}.
\end{equation}
For the reference part with known solution and the generalized one using of (\ref{a11}) to be solved:
\begin{equation} \label{a14}
{{\left( \frac{{\dot{a_r}}}{a_r} \right)}^{2}}=\frac{8\pi {{G}_{N}}}{3}\rho_{r}=
\frac{\frac{8\pi {{G}_{N}}}{3}{{\rho }_{0}}{{a}^{-3(1+w )}}}{\left( 1+\frac{\lambda b{{\rho }_{0}}}{(1+w )}\left( 1-{{a}^{-3(1+w )}} \right) \right)}.
\end{equation}
After some calculations (Appendix B), the final expression for the scale factor will be obtained as:
$$a_{r}(t)=\left( 1-\frac{3b{{\rho }_{0}}{{n}^{2}}}{4} \right){{\left( \frac{t}{{{t}_{0}}} \right)}^{n}}$$
\begin{equation} \label{a23}
+\frac{6b{{\rho }_{0}}{{n}^{2}}}{4}{{\left( \frac{t}{{{t}_{0}}} \right)}^{n-1}}-\frac{3b{{\rho }_{0}}{{n}^{2}}}{4}{{\left( \frac{t}{{{t}_{0}}} \right)}^{n-2}},
\end{equation}
where we have,
\begin{equation} \label{a201}
n=\frac{2}{3(1+w )}.
 \end{equation}
We may note that if we started (\ref{hard7}), we would get the same expression. We can write equation $(\ref{a23})$ as,
\begin{equation}
\label{hard111}
a_{r}(t)=( 1-\delta){{\left( \frac{t}{{{t}_{0}}} \right)}^{n}}+2\delta{{\left(\frac{t}{{{t}_{0}}} \right)}^{n-1}}-\delta{{\left( \frac{t}{{{t}_{0}}} \right)}^{n-2}},
\end{equation}
where we take $\delta=\frac{3b{{\rho }_{0}}{{n}^{2}}}{4}=\frac{b{\rho}_{0}}{3(1+w)^2}= \frac{n{\alpha}}{2}$.  As can be seen, we have in the limit of $t\rightarrow t_0$, $a\rightarrow a_0$. For the case of $\alpha=0$, we have the conventional form of the scale factor, i.e. $a(t)={{\left( \frac{t}{{{t}_{0}}} \right)}^{n}}$. Also, for $(t/t_0)\rightarrow{0}$, we should have $a\rightarrow{0}$. \\
\section{Some Discussions and Possible Results}
As expected, compared to the conventional relation $(\ref{hard6})$, our relation for the energy density $(\ref{hard71})$, is divided into two different parts; the first part is essentially the same as standard one, except that now a certain fraction of it has been subtracted. The second part has exactly the same fraction coming from the second term of the equation of state $(\ref{hard1})$  containing the equilibrium constant. We can state that the cosmic gas in our scenario includes two different parts, the first part of (\ref{hard71})  is related to under-dense regions with lower pressure whereas the second part is related to regions with higher density and pressure. In the limit of $\alpha = 0$, $(\ref{hard71})$ returns to the conventional result $(\ref{hard6})$. However, for the present cold universe containing clustering filaments in the gas with $\alpha \neq 0$, the correction term gets important and is applicable for the regions where galaxies (or simultaneously voids) merge together to form clusters. \\
Since, we will get the standard solution at the limit of $\alpha=0$, $\alpha$ can be regarded as a correction factor to relation $(\ref{hard6})$. With the continuation of the expansion of the cosmos and with the increasing of the alpha factor, the second term will account for a larger contribution of the cosmic density. Note that the $\alpha$ values are related to the intensity of the gravitational merging among cosmic objects. As seen the $\alpha$ values depend on three parameters $\rho_0$, $b$, and $w$. $\rho_0$ is known, hence the values of $b$ and $w$ should be adjusted to give the suitable $\alpha$ values.\\
\subsection{Energy Density and Deceleration Parameter for a Void-dominant Cosmic Fluid}
According to our proposal presented in Appendix B, the cosmic fluid at present time (cosmic web) contains mergings of several vast voids that are located in the \emph{void-dominant} phase. Therefore the negative pressure for such a fluid seems logical and even obvious (see Appendix B). \\
Fortunately, cosmic data is also looking for such a negative pressure for the cosmic gas. The observational data show  that parameter $w$ has a very narrow range around the pure vacuum equation of state $w=-1$, with more likelihood to the side of $w\lesssim -1$ \cite{c11}. So, for this range of $w$, we have $\alpha > 0$. The positive sign for the energy density in second term of Eq.$(\ref{hard71})$, can have the meaning of the gravitational clustering of the galaxies, but with the negative sign in the first term of Eq.$(\ref{hard71})$ that can mean as repulsive gravitational force (expansion). This additional repulsion force in the inter galactic gas (inside the voids) causes extra-acceleration in the large scale universe. On the other hand, for the range of $w\gtrsim -1$, we have $\alpha < 0$. That is, the negative sign for the energy density in second term of Eq.$(\ref{hard71})$, could have the meaning of gravitational repulsion due to integration of voids and the formation of giant voids to form the shape of the cosmic web (see Fig. 2). As a key point, we can state that the merging of the cosmic voids at their's surfaces produce \emph{both} of clustering (positive pressure) in filaments at small scales and accelerating (negative pressure) at large scales, simultaneously. In a very similar way to what is found in the soap bubbles \cite{c202} or overflowing process in the boiling milk \cite{c23}. We believe that in the Farnes model \cite{c24}, the negative mass corresponds to voids or equivalently bubbles. \\
 \begin{figure}[t]
    \centering
    \includegraphics[width=3 in]{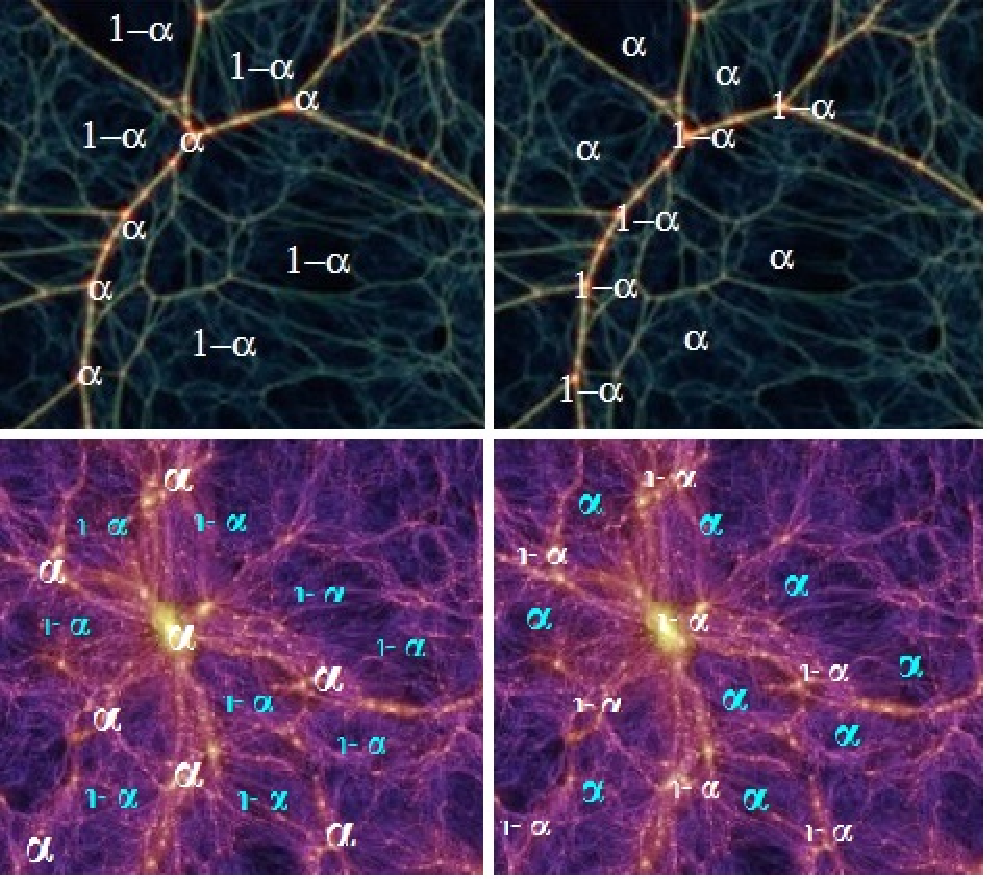}
\caption{The cosmic web: $\alpha>0$ (left) for the higher-dense regions (galaxy clusters) \emph{versus} $\alpha<0$ (right) for the lower-dense regions (voids).}
          \label{Fig2}
\end{figure}
  In Fig. 3 the energy density $\rho/\rho_0$ is plotted in the range of $(-2 <w < 0)$ for the PCF model and RCF models with $\alpha=0.3$ and $\alpha=-0.3$. As shown in the figure 3, except for the particular value $w=-1$, the energy density of the universe at large scale with the assumption of gravitational merging, is \emph{always lower} than the energy density of cosmic gas with non-merging objects.\\

\begin{figure}[h]
    \centering
    \includegraphics[width=2.3 in]{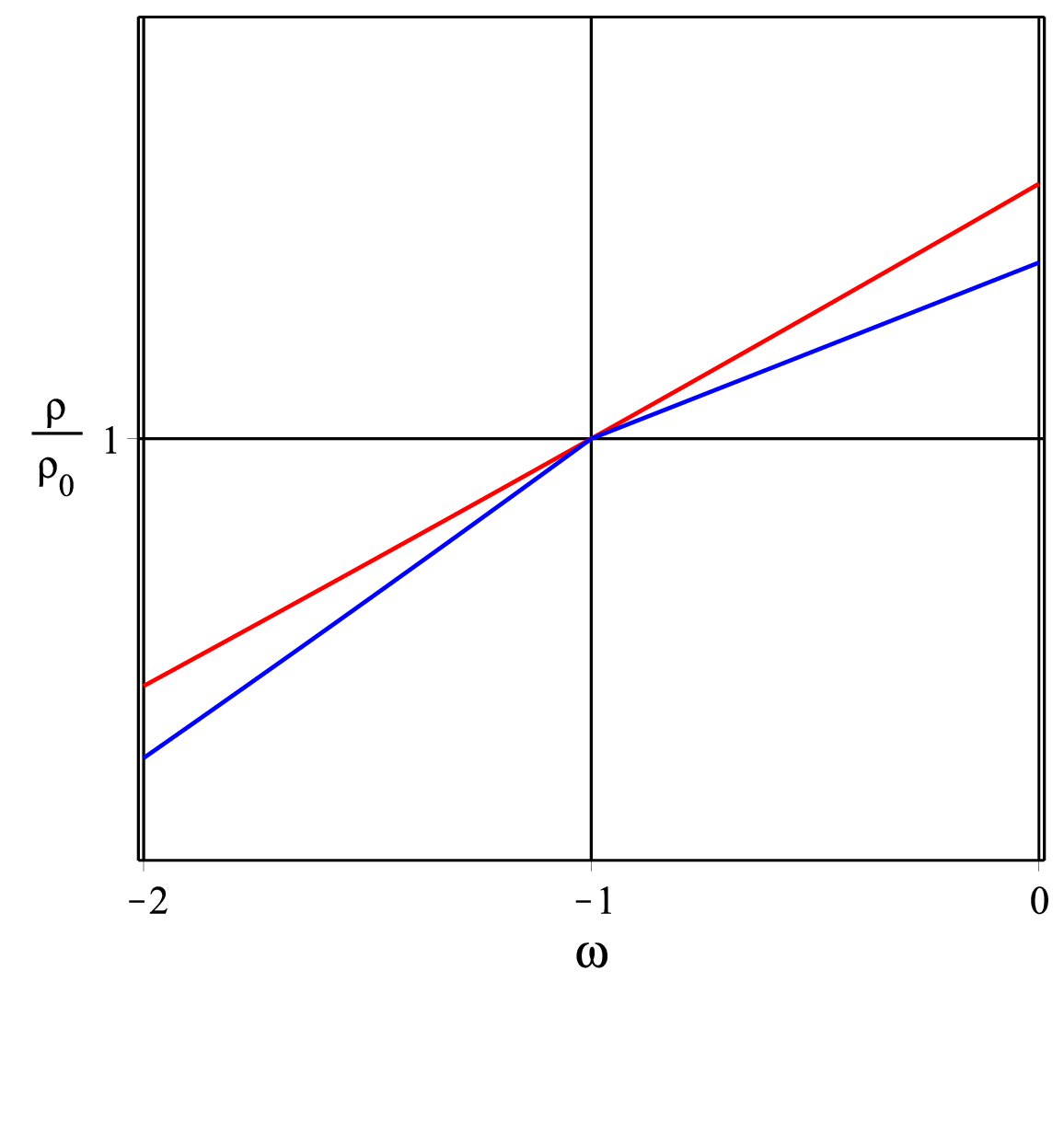}
    \caption{Energy density of the universe is plotted in terms of $w$  for the PCF model (red) and RCF models (blue) for $\alpha=0.3$ (left) and $\alpha=-0.3$ (right) for $a=0.99$.}
          \label{Fig3}
\end{figure}
The deceleration parameter $q$, in cosmology is defined as \cite{c4, c7},
\begin{equation} \label{hard97}
q=-\frac{\ddot{a}}{aH^{2}}=-\frac{a\ddot{a}}{(\dot{a})^2}|_{t=t_0}
 \end{equation}
 The larger value of $q$ with negative sign indicates more rapid acceleration. It is a way to quantify the accelerated expanding of present universe at $t=t_0$. After simple calculations we will obtain the Hubble parameter at $t_0 = 1$ as,
\begin{equation} \label{hard979}
H_{r}=H_{p}=n,
\end{equation}
and deceleration parameter as,
\begin{equation} \label{hard981}
q_{r}=-1+\frac{1}{n}+\frac{\alpha}{n}.
 \end{equation}
 \\
 From the fact that $b<0$ for the merging process, the sign of $\frac{\alpha}{n}=\frac{3b\rho_0}{2}$ is always negative. So, we can rewrite
\begin{equation} \label{hard982}
q_{r}=q_{p}+\frac{\alpha}{n}.
 \end{equation}
and consequently \emph{we have $q_{r}$ larger than $q_{p}$ with negative sign}.
\begin{figure}[h]
    \centering
    \includegraphics[width=2.3 in]{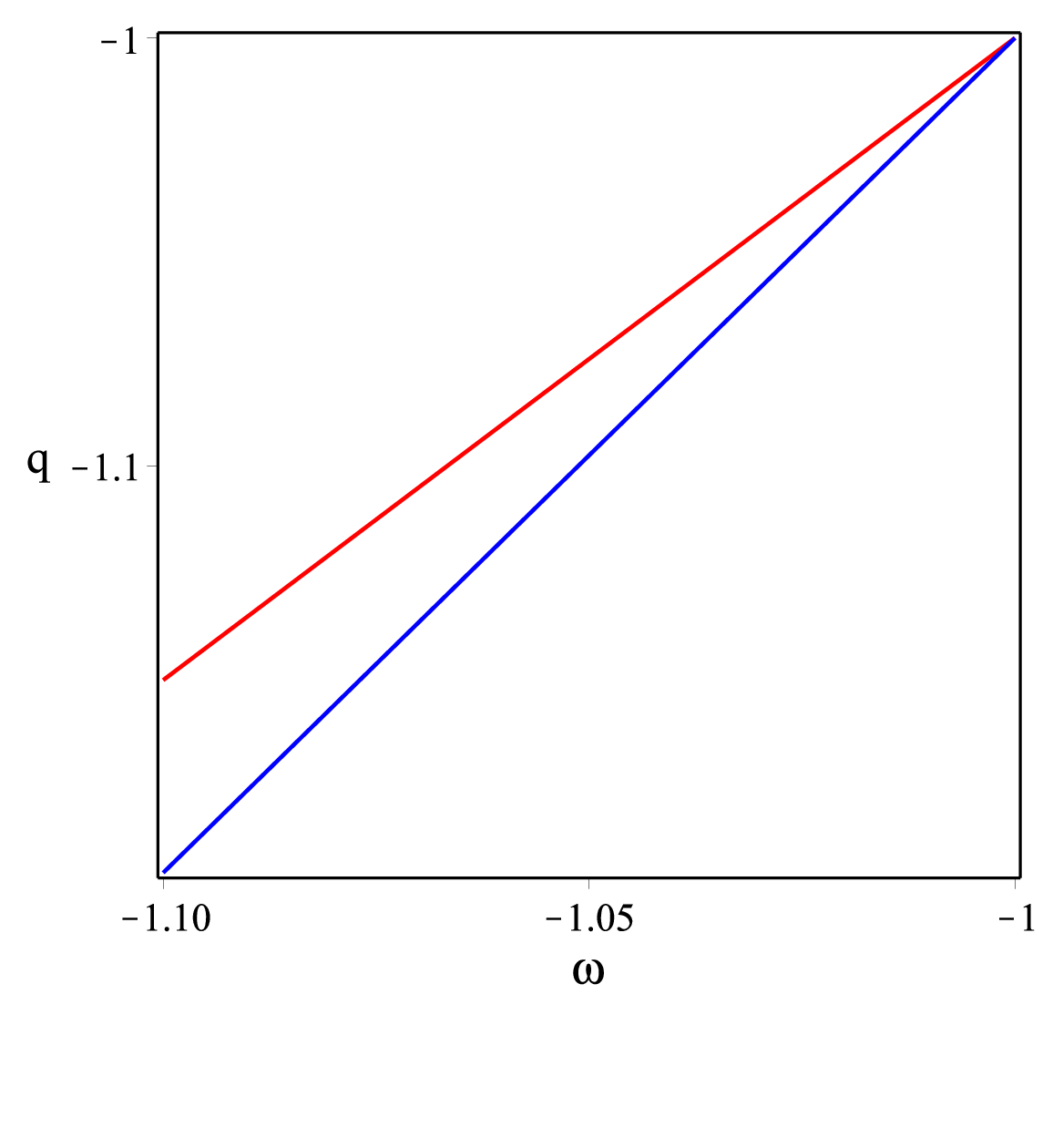}
    \caption{Deceleration parameter is plotted in terms of $w$ for
    the PCF model (red) and RCF models for $\alpha=+0.3$ (blue).}
          \label{Fig4}
\end{figure}
\begin{figure}[h]
    \centering
    \includegraphics[width=2.3 in]{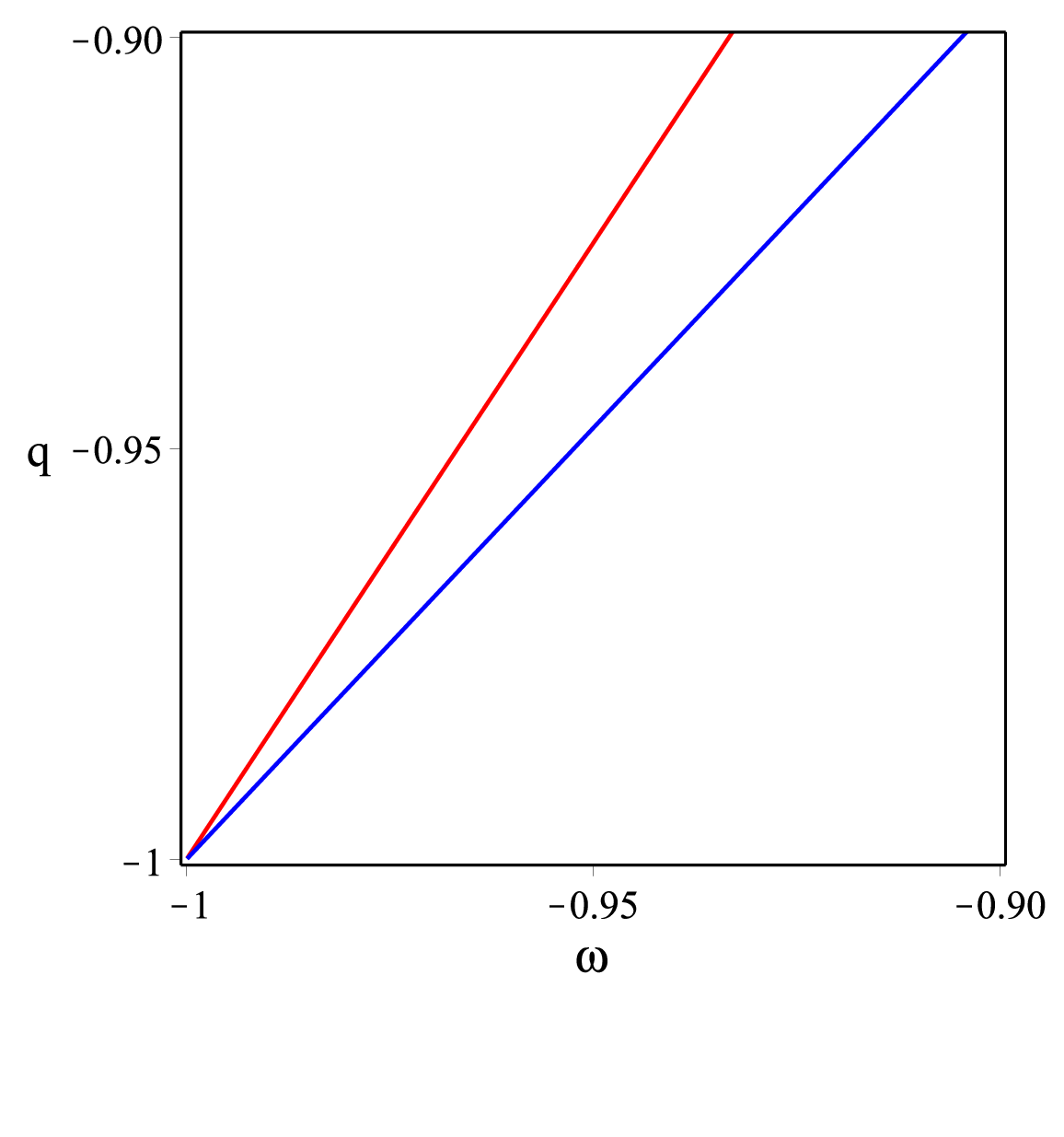}
\caption{Deceleration parameter is plotted in terms of $w$ for
    the PCF model (red) and RCF models for $\alpha=-0.3$ (blue).}
          \label{Fig5}
\end{figure}

Let us compare the deceleration parameters of both PCF and RCF models in terms of parameter $w$. In Figs. 4 and 5 the deceleration parameters $q$ is plotted versus $w$ in the range of $(-1.1<w <-1)$ and  $(-1<w <-0.9)$, respectively. As seen, the plots cross each other at a particular value $w=-1$.  Since the observed values for parameter $w$ for type Ia supernovae, the equation of state of dark energy is constrained to $w = -1.006\pm 0.045 $ \cite{c11}, by consideration $w = -1.006$, we obtain deceleration parameter for PCF model $q = -1.009 $ , and for RCF model, $q = -1.012 $. The significant result is that the acceleration for the both phantom $(w\lesssim -1, \alpha > 0)$ and non-phantom $(w\gtrsim -1, \alpha < 0)$  phases of the cosmic fluid in the RCF model is larger than PCF model (The blue lines in the Figs. 4 and 5 situated under red line).\\
\section{Summary and Conclusions}
Merging of cosmic objects at small scales is regarded as a possible source for explaining the extra-acceleration of the universe at large scale. Inspired by association process, merging of cosmic objects introduces a correction term in the perfect fluid equation of state in the form of $P=w\rho+b\rho^2$. Alternative relations for the energy density and scale factor evolution in the FRLW framework are obtained that coincide with the conventional results in the standard limit. Invoking the observational constraint for the equation of state parameter $w$, we show that the merging of cosmic objects such as galaxies and voids simultaneously will act as a possible source of extra-accelerating in the universe at large scale. \\
Similar to standard model of cosmology that consider perfect fluid in the vacuum, radiation and matter phases; we have proposed that the real cosmic fluid composed of two phases, voids and galaxies clusters for the evolution of the present universe. According to energy density relation $(\ref{hard71})$ and the results obtained for the Hubble and deceleration parameters, it seems that the extra-acceleration at large scale of the universe coming from the integration and clustering of the cosmic objects in the small scales of the universe. Neither voids nor galaxies
clustering/merging are new ideas. When considered individually, neither idea can explain acceleration at large scales. Our model in this paper has reinvoked these two concepts in the unify form in the real cosmic fluid and combined them together. As a result, an alternative relation for the energy density includes over and under-dense regions is obtained.  Also, by analogy with bubbles, the under-dense regions (voids) in the cosmic fluid is shown to provide the needed negative pressure. It naturally implies a beautiful symmetry, in which all physical systems are polarised into positive and negative states. A polarised real cosmic fluid that contains both clusters and voids with positive and negative pressures can literally bring balance to the cosmic system.\\
As a key result, we can conclude that the merging of the cosmic voids at their's surfaces produce \emph{both} of clustering (with positive pressure) at small scales and accelerating (with negative pressure) at large scales, simultaneously. In a very similar way to what is found in the soap bubbles or overflowing process in the boiling milk. We think that the source of this negative pressure (or repulsive gravity) is the surface tension at the surface of the voids. While we know that the surfaces of these bubbles is forming due to the gravitational clustering of the galaxies, the main force behind clustering is the attractive gravity. \emph{So, the attractive gravity at small scales will in turn produce the repulsive gravity at large scales.}\\
In this work we have presented the basic ideas and mathematical derivations of the model. Based on the RCF scenario, it seems that the model can provide predictions for dark matter in addition to dark energy. The work on this subject is underway. Also, some later works will naturally be devoted to the observational and some more interesting consequences from void-based cosmology.
 \vspace*{4mm}

\begin{acknowledgments}
  EY would like to acknowledge David H. Lyth, H. Firouzjahi, Sh. Baghram, S. Tavasoli, M. Mohsenzadeh and M. V. Takook for their help in improving the manuscript. EY would like to thank school of Astronomy at IPM for the material and spiritual support during preparing initial version of this research. This work has been supported by the Islamic Azad University, Ayatollah Amoli Branch, Amol, Iran.

\end{acknowledgments}
\appendix

\section{Derivation of the Scale Factor}
In combination with Eq.$(\ref{hard69})$, Eq. (\ref{a14}) becomes:
\begin{equation} \label{a15}
{{\left( \frac{{{{\dot{a}}}}+\lambda {{{\dot{a}}}_{1}}}{{{a}}+\lambda {{a}_{1}}} \right)}^{2}}=
\frac{\frac{8\pi {{G}_{N}}}{3}{{\rho }_{0}}{{({{a}_{0}}+\lambda {{a}_{1}})}^{-3(1+w )}}}{\left( 1+\frac{\lambda b{{\rho }_{0}}}{(1+w )}\left( 1-{{({{a}}+\lambda {{a}_{1}})}^{-3(1+w )}} \right) \right)}
\end{equation}
Or after taking out the reference system from both sides:
\begin{equation} \label{a16}
{{\left( \frac{{{{\dot{a}}}}}{{{a}}} \right)}^{2}}{{\left( \frac{1+\lambda {{{\dot{a}}}_{1}}/{{{\dot{a}}}}}{1+\lambda {{a}_{1}}/{{a}}} \right)}^{2}}=
\frac{\frac{8\pi {{G}_{N}}}{3}{{\rho }_{0}}a^{-3(1+w )}{{(1+\lambda {{a}_{1}}/{{a}})}^{-3(1+w )}}}{\left( 1+\frac{\lambda b{{\rho }_{0}}}{(1+w )}\left( 1-a^{-3(1+w )}{{(1+\lambda {{a}_{1}}/{{a}})}^{-3(1+w )}} \right) \right)}
\end{equation}
Removing the reference parts we are left with:
\begin{equation} \label{a17}
{{\left( \frac{1+\lambda {{{\dot{a}}}_{1}}/{{{\dot{a}}}}}{1+\lambda {{a}_{1}}/{{a}}} \right)}^{2}}=
\frac{{{(1+\lambda {{a}_{1}}/{{a}})}^{-3(1+w )}}}{\left( 1+\frac{\lambda b{{\rho }_{0}}}{(1+w )}\left( 1-a^{-3(1+w )}{{(1+\lambda {{a}_{1}}/{{a}})}^{-3(1+w )}} \right) \right)}.
\end{equation}
In order to make this equation be solvable we will expand both sides in terms of $\lambda$ up to first order:
$$1+2\left( \frac{{{{\dot{a}}}_{1}}}{{{{\dot{a}}}}}-\frac{{{a}_{1}}}{{{a}}} \right)\lambda +...=$$
\begin{equation} \label{a18}
1+\left( -\frac{b{{\rho }_{0}}(1-a^{-3(1+w )})}{1+w }-\frac{3(1+w ){{a}_{1}}}{{{a}}} \right)\lambda +....
\end{equation}
Fortunately, this will lead to a linear differential equation for finding $a_1$ in terms of known $a$:
\begin{equation} \label{a19}
\frac{2{{{\dot{a}}}_{1}}}{{{{\dot{a}}}}}-\frac{2{{a}_{1}}}{{{a}}}=-\frac{3(1+w ){{a}_{1}}}{{{a}}}-\frac{b{{\rho }_{0}}(1-a^{-3(1+w )})}{1+w },
\end{equation}
which may be cast in a simpler form:
\begin{equation} \label{a20}
{{\dot{a}}_{1}}+\left( \frac{3(1+w )}{2}-1 \right)\left( \frac{{{{\dot{a}}}}}{{{a}}} \right){{a}_{1}}=-\frac{b{{\rho }_{0}}{{{\dot{a}}}}(1-a^{-3(1+w )})}{2(1+w )}.
\end{equation}
Let's invoke the reference system solution. It reads as ${{a}{(t)}}={{\left( \frac{t}{{{t}_{0}}} \right)}^{n}},$
satisfying naturally the condition at $t=t_0$, $a=1$. Also, putting it in Eq.$(\ref{a13})$ leads to $(\ref{a201})$, and $\frac{{{{\dot{a}}}_{t}}}{{{a}}}=\frac{n}{t}$ needed later. Plugging them into Eq.$(\ref{a20})$ will transform it into:
\begin{equation} \label{a21}
{{\dot{a}}_{1}}+\left( \frac{1-n}{t} \right){{a}_{1}}=-\frac{3b{{\rho }_{0}}{{n}^{2}}}{4t}\left( {{\left( \frac{t}{{{t}_{0}}} \right)}^{n}}-{{\left( \frac{t}{{{t}_{0}}} \right)}^{n-2}} \right).
\end{equation}
This is a simple first order differential equation with the following solution:
\begin{equation} \label{a22}
{{a}_{1}}=-\frac{3b{{\rho }_{0}}{{n}^{2}}}{4}\left\{ {{\left( \frac{t}{{{t}_{0}}} \right)}^{n}}+{{\left( \frac{t}{{{t}_{0}}} \right)}^{n-2}} \right\}+\frac{6b{{\rho }_{0}}{{n}^{2}}}{4}{{\left( \frac{t}{{{t}_{0}}} \right)}^{n-1}}.
\end{equation}
Where we have used the condition at $t=t_0$, $a_1=0$, and hence the final expression for the scale factor will be obtained as $(\ref{a23})$.
\section{On the Origin of the Negative Pressure}
In this Appendix we will discuss on the signature of the present cosmic web pressure. Let us consider a small droplet with spherical shape in equilibrium with its vapor. The droplet and the saturated vapor are enclosed in a constant temperature rigid container shown in Fig. \ref{Fig6}.\\
\begin{figure}[h]
\centering
\includegraphics[width=3in]{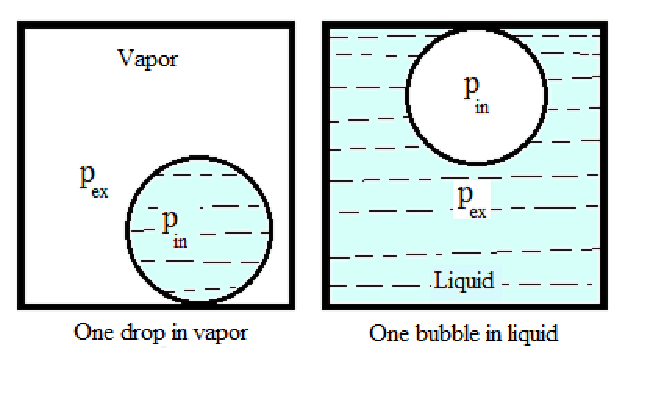}
\caption{Drop in vapor (left) v.s. Bubble in liquid (right)\cite{c21}.}
\label{Fig6}
\end{figure}
As shown in \cite{c260}, the equilibrium pressure within the drop is as follows:
\begin{equation}
\label{hard711}
P_{in(drop)}=P_{ex(drop)}+\frac{2\gamma}{\bar{r}}.
\end{equation}
Here, $\gamma$ represents the surface tension for droplet with radius $\bar{r}$. In case of a spherical bubble interfaced with liquid, the liquid surface is not convex but concave i.e., its curvature is not positive but negative \cite{c260}. As a consequence, the equilibrium pressure inside a bubble is:
\begin{equation}
\label{hard712}
P_{in(bubble)}=P_{ex(bubble)}-\frac{2\gamma}{\bar{r}}.
\end{equation}

Inspired by equation $(\ref{hard71})$, we can consider the fluid of the present universe at large scales can be regarded as a two-phase mixed fluid including drops (over-dense regions) and bubbles (under-dense regions). Galaxy clusters behave like drops and voids like bubbles. They are connected and merged together on their's surfaces. Since zero pressure is assumed for the galaxies and their's clusters in present universe, i.e. $P_{ex(bubble)}=P_{galaxies}\approx 0$, as an important result we conclude from $(\ref{hard712})$ that the total pressure of the cosmic web at large scale can be negative.
\begin{equation}
\label{hard714}
P_{web}\approx P_{voids}< P_{galaxies} \approx 0.
\end{equation}
\emph{Since the dominant volume of the cosmic web is formed by voids, based the above picture, the negative pressure is inevitable at large scale. That is, the cosmic gas is considered as composed of so many bubbles or voids, each of which have pressure less than the clusters pressure, and hence the total pressure of the universe would become negative. In other words our universe at local small scales is matter-dominated and at global cosmic scales is void-dominated.}

\end{document}